\documentclass[twocolumn]{aastex63}

\newcommand{\oi}{O\,{\sc i}}
\newcommand{\siiv}{Si\,{\sc iv}}
\newcommand{\nv}{N\,{\sc v}}
\newcommand{\civ}{C\,{\sc iv}}
\newcommand{\ciii}{C\,{\sc iii]}}
\newcommand{\mgii}{Mg\,{\sc ii}}
\newcommand{\cii}{[C\,{\sc ii}]}

\usepackage{amsmath}

\usepackage{physics}

\shorttitle{A Close Quasar Pair Candidate at $z=5.66$}
\shortauthors{Yue et al.}

\begin{document}

\title{A Candidate Kiloparsec-scale Quasar Pair at $z=5.66$}

\correspondingauthor{Minghao Yue}
\email{yuemh@email.arizona.edu}

\author[0000-0002-5367-8021]{Minghao Yue}
\affiliation{Steward Observatory, University of Arizona, 933 North Cherry Avenue, Tucson, AZ 85721, USA}

\author[0000-0003-3310-0131]{Xiaohui Fan}
\affiliation{Steward Observatory, University of Arizona, 933 North Cherry Avenue, Tucson, AZ 85721, USA}

\author[0000-0001-5287-4242]{Jinyi Yang}
\altaffiliation{Strittmatter Fellow}
\affil{Steward Observatory, University of Arizona, 933 North Cherry Avenue, Tucson, AZ 85721, USA}

\author[0000-0002-7633-431X]{Feige Wang}
\altaffiliation{NHFP Hubble Fellow}
\affiliation{Steward Observatory, University of Arizona, 933 North Cherry Avenue, Tucson, AZ 85721, USA}



\begin{abstract}

We report the discovery of a close quasar pair candidate at $z=5.66$, J2037--4537.
J2037--4537 is resolved into two quasar images at the same redshift in ground-based observations. 
Followup spectroscopy shows significant differences in both the continuum slopes and emission line properties of the two images. 
The two quasar images have a projected separation of $1\farcs24$ ($7.3\text{~kpc}$ at $z=5.66$)
and a redshift difference of $\Delta z\lesssim0.01$.
High-resolution images taken by {\em Hubble Space Telescope} do not detect the foreground lensing galaxy.
The observational features of J2037--4537 strongly disfavor the lensing hypothesis.
If J2037--4537 is a physical quasar pair, it
indicates a quasar clustering signal of $\sim10^5$ at a separation of $\sim10$ proper kpc (pkpc),
and gives the first observational constraint on the pair fraction of $z>5$ quasars,
$f_\text{pair}(r<30\text{~pkpc})>0.3\%$.
The properties of J2037--4537 are consistent with those of merger-triggered quasar pairs in hydrodynamical simulations of galaxy mergers. 
\end{abstract}





\section{Introduction} \label{sec:intro}

Galaxy mergers are natural consequences of hierarchical structure formation of the universe \citep[e.g.,][]{cole00}.
It has been proposed that galaxy mergers can trigger powerful quasar activity
which regulates the star formation of their host galaxies by significant feedback \citep[e.g.,][]{dm05}.
In some rare cases, the supermassive black holes (SMBHs) in both progenitor galaxies 
are ignited by the merging event, forming a close pair of quasars \citep[for a recent review, see][]{dr19}.
High-resolution hydrodynamical simulations suggest that close quasar pairs correspond to a special, short-lived
phase of galaxy mergers, which most frequently appear when the SMBHs have a mass ratio close to one and
a separation smaller than 10 kpc \citep[e.g.,][]{capelo15, capelo17}.
Quasar pairs are also predicted in cosmological simulations \citep[e.g.,][]{steinborn16,volonteri16, rg19}, 
and their statistics (e.g., the fraction of pairs among all quasars) 
constrain the evolution of SMBHs and their host galaxies.
In addition, close quasar pairs trace overdensities \citep[e.g.,][]{onoue18}, 
and are unique probes of the small-scale structure of intergalactic medium \citep[e.g.,][]{rorai17}.

Searches of quasar pairs have been carried out using wide-area optical and infrared (IR) sky surveys 
\citep[e.g.,][]{hennawi10, silverman20}.
These studies have discovered several tens of close quasar pairs 
(with projected separation $\Delta d\lesssim10$ kpc) out to $z\sim3$ \citep[e.g.,][]{shen21, tang2021}.
Due to the rapid decline of quasar number densities \citep[e.g.,][]{kulkarni19} and the relatively poor physical resolution for most observations,
it becomes difficult to identify close quasar pairs at $z\gtrsim3$.
Finding high-redshift quasar pairs is critical to the understanding of the evolution and environment of 
high-redshift quasars and galaxy mergers.

In this letter, we report a close quasar pair candidate 
at redshift $z=5.66$ 
with a projected proper distance of $\Delta d=7.3$ kpc, J2037--4537.
Following \citet{hennawi10}, in this Letter, we define a quasar pair as two quasars with 
 $\Delta d<1\text{ Mpc}$ and a radial velocity difference of
$\Delta v<2000\text{ km s}^{-1}$. We further define close quasar pairs to be those with  
$\Delta d<10\text{ kpc}$. 
The most distant quasar pair previously known is at $z=5.1$ with $\Delta d=135$ kpc \citep{mcgreer16},
and for close quasar pairs, the highest-redshift one is 
at $z=3.1$ with $\Delta d=7.4\text{ kpc}$ \citep{tang2021}. 
J2037--4537 significantly extends the frontier of studies of quasar pairs.

This letter is organized as follows.
We describe the ground-based observations and the {\em Hubble Space Telescope} 
({\em HST}) imaging in Section \ref{sec:data}.
We discuss the interpretation of the data in Section \ref{sec:discussion} and summarize in Section \ref{sec:sum}.
We use a flat $\Lambda$CDM cosmology with $H_0=70\text{ km s}^{-1} \text{ Mpc}^{-1}$ and $\Omega_M=0.3$.

\section{Data and Analysis} \label{sec:data}


\subsection{Ground-Based Imaging and Spectroscopy}

\begin{figure*}
\centering
\includegraphics[trim=0.5cm 0cm 0.5cm 0cm,clip,width=0.16\linewidth]{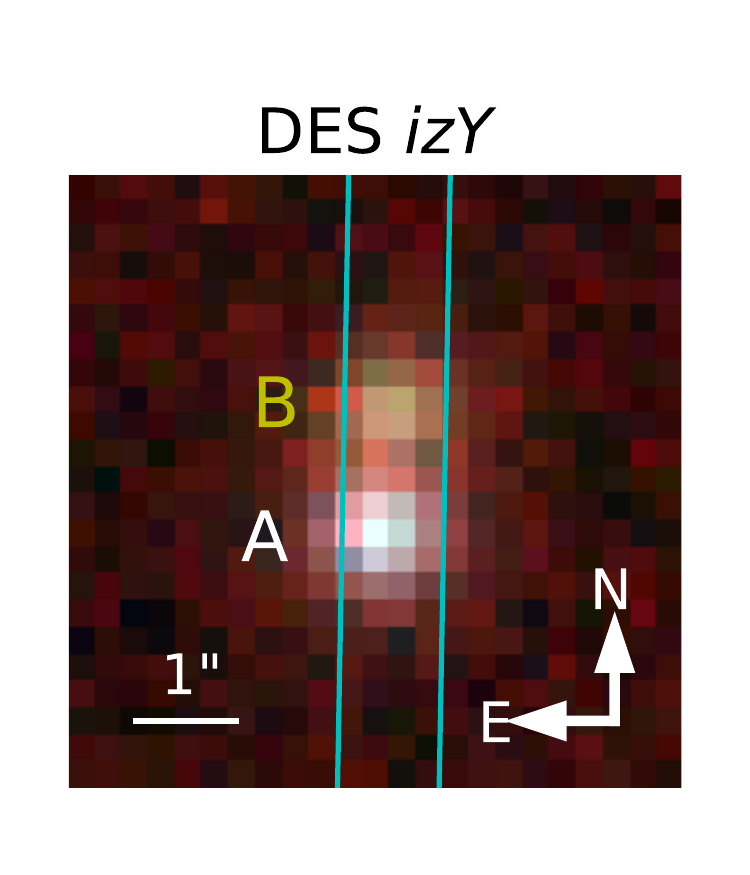}
\includegraphics[trim=3.3cm 0cm 3.3cm 0cm,clip,width=0.7\linewidth]{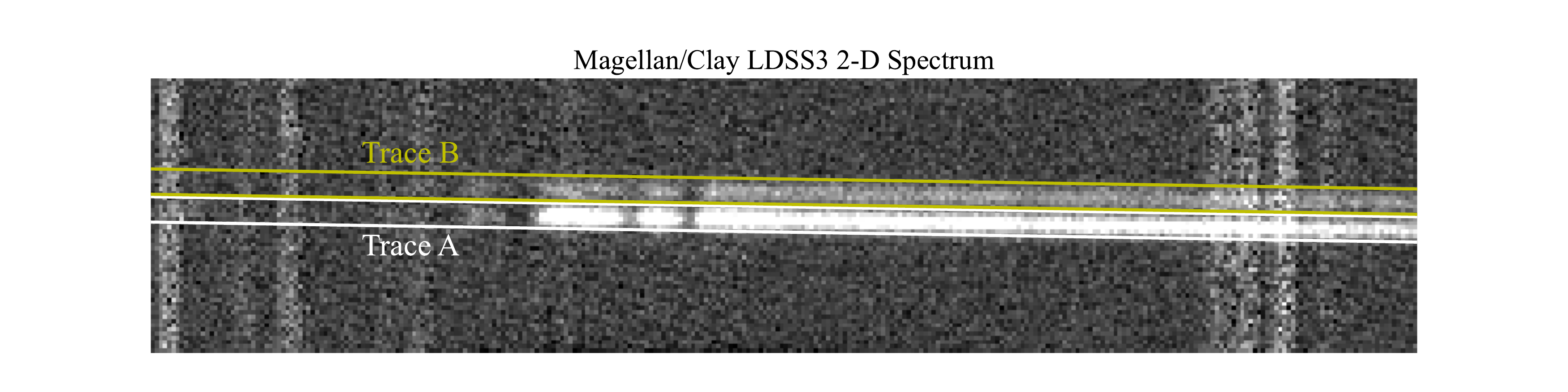}\\
\centering
\includegraphics[width=0.95\linewidth]{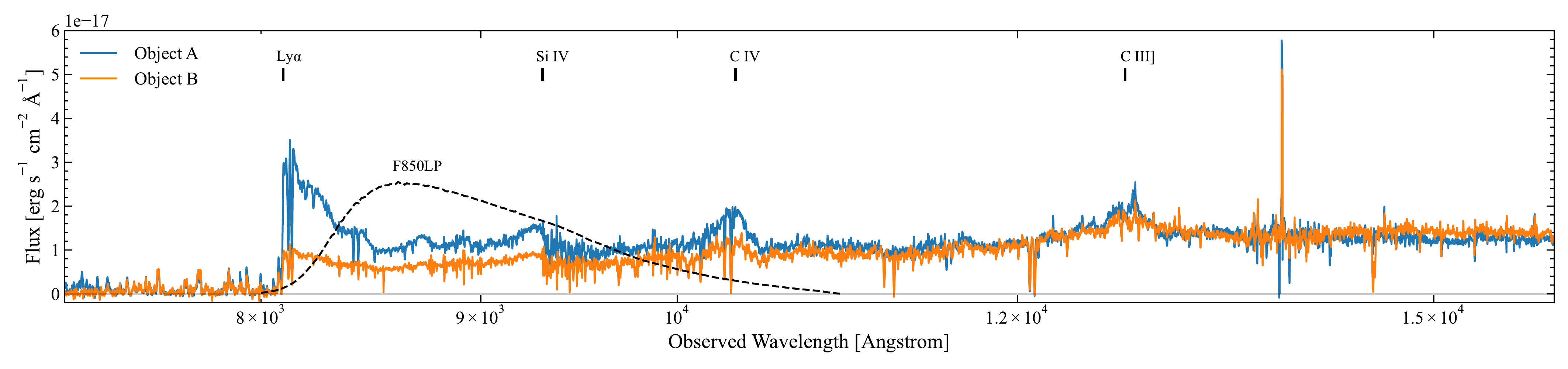}
\includegraphics[width=0.95\linewidth]{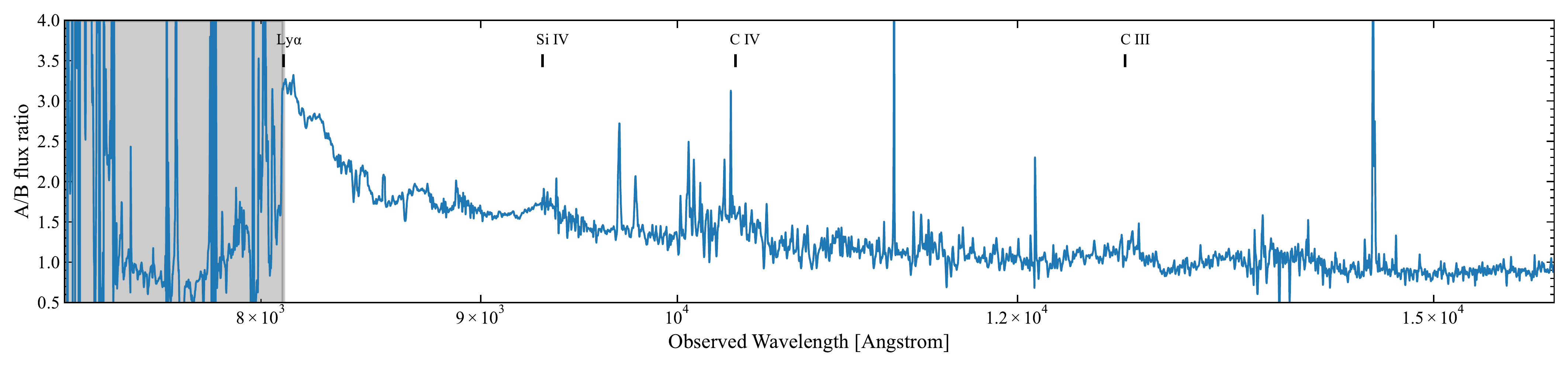}
\caption{Ground-based imaging and spectroscopy of J2037--4537. {\em Upper Left}: the DES $izY$ images shown in RGB format.
The PSF FWHM of the image is $\sim0\farcs9$. The fainter component (object B) is redder than the brighter one (object A).
The two objects are separated by $1\farcs24$. The cyan lines mark the slit position in the spectroscopy.
{\em Upper Right}: the 2-D spectrum taken by Magellan/Clay LDSS-3. 
The two objects are well-resolved under a seeing of $\sim0\farcs7$, allowing us to extract the two traces accurately.
The white and yellow lines mark the apertures used to extract the 1-D spectrum.
The Magellan/Baade FIRE spectrum has a better seeing of $\sim0\farcs6$.
{\em Middle Panel}: The extracted 1-D spectrum of the two objects, which is a combination of the
Magellan/Clay LDSS-3 spectra $(\lambda<1\mu\text{m})$ and the Magellan/Baade FIRE spectra $(\lambda>1\mu\text{m})$.
Both objects exhibit features of a $z=5.66$ quasar, and object B is redder than object A.
The dashed line shows the F850LP filter used in {\em HST} ACS/WFC imaging.
{\em Lower Panel}: The flux ratio between the two objects. 
The shaded area marks wavelengths bluer than the Ly$\alpha$ break, where the quasars have little flux
due to the IGM absorption.
There are complex features around the emission lines,
which suggest that the two objects have different line profiles, and disfavor the strong lensing hypothesis.}
\label{fig:ground}
\end{figure*}

J2037--4537 was discovered in our on-going survey 
for high-redshift gravitationally lensed quasars (Yue et al. in prep)
in the Dark Energy Survey \citep[DES, e.g.,][]{des} field.
It is not detected in the DES $g-$band, and has the DES $r=22.8$, $i=20.4$, $z=19.6$ and $Y=19.5$, 
exhibiting typical colors of a quasar at redshift $z\gtrsim5.5$ \citep[e.g.,][]{yang19}. 
In the DES images (Figure \ref{fig:ground}, upper left), J2037--4537 appears to be two point sources separated by $1\farcs24$, where the fainter one (object B) is redder than the brighter one (object A).
Table \ref{tbl:properties} lists the photometric properties of the two objects.
The magnitudes are obtained by fitting the images as two point sources using {\em galfit} \citep{galfit}, 
where the point spread function (PSF) models are built by the IRAF task {\texttt{psf}}.

The optical spectrum of J2037--4537, taken by the Low Dispersion Survey Spectrograph (LDSS-3) on Magellan/Clay telescope,
reveals that both objects are quasars at redshift $z=5.66$ (Figure \ref{fig:ground}).
We use the VPH-Red grism and the $1\farcs0$ slit which deliver a resolution of $R=1350$. 
The spectrum is reduced with the standard IRAF pipeline.
The two objects are clearly resolved in the spectrum and have the same Ly$\alpha$-break wavelength,
suggesting a redshift difference $\Delta z\lesssim0.01$.
Object B has a redder spectral energy distribution (SED) shape than object A, which agrees with the broad-band photometries.
We also obtained near-IR spectra of the two objects using 
the Folded-port InfraRed Echelle (FIRE) on Magellan/Baade telescope.
The spectra are reduced with {\texttt{PypeIt}} \citep{pypeit}.

J2037--4537 could either be 
a physical quasar pair, or two images of a strongly lensed quasar. 
Both the image and the spectra do not show signs of a third object (e.g., a foreground lensing galaxy).
This is further confirmed by the {\em HST} image (Section \ref{subsec:hst}).
In addition, the two objects have different emission line profiles.
The difference is the most obvious for the {\ciii} line and can also be seen in the A/B flux ratio 
(Figure \ref{fig:ground}, lower panel).
These features disfavor the strong-lensing scenario and 
indicate that J2037--4537 is a physical quasar pair.

\begin{figure}
    \centering
    \includegraphics[width=0.98\columnwidth]{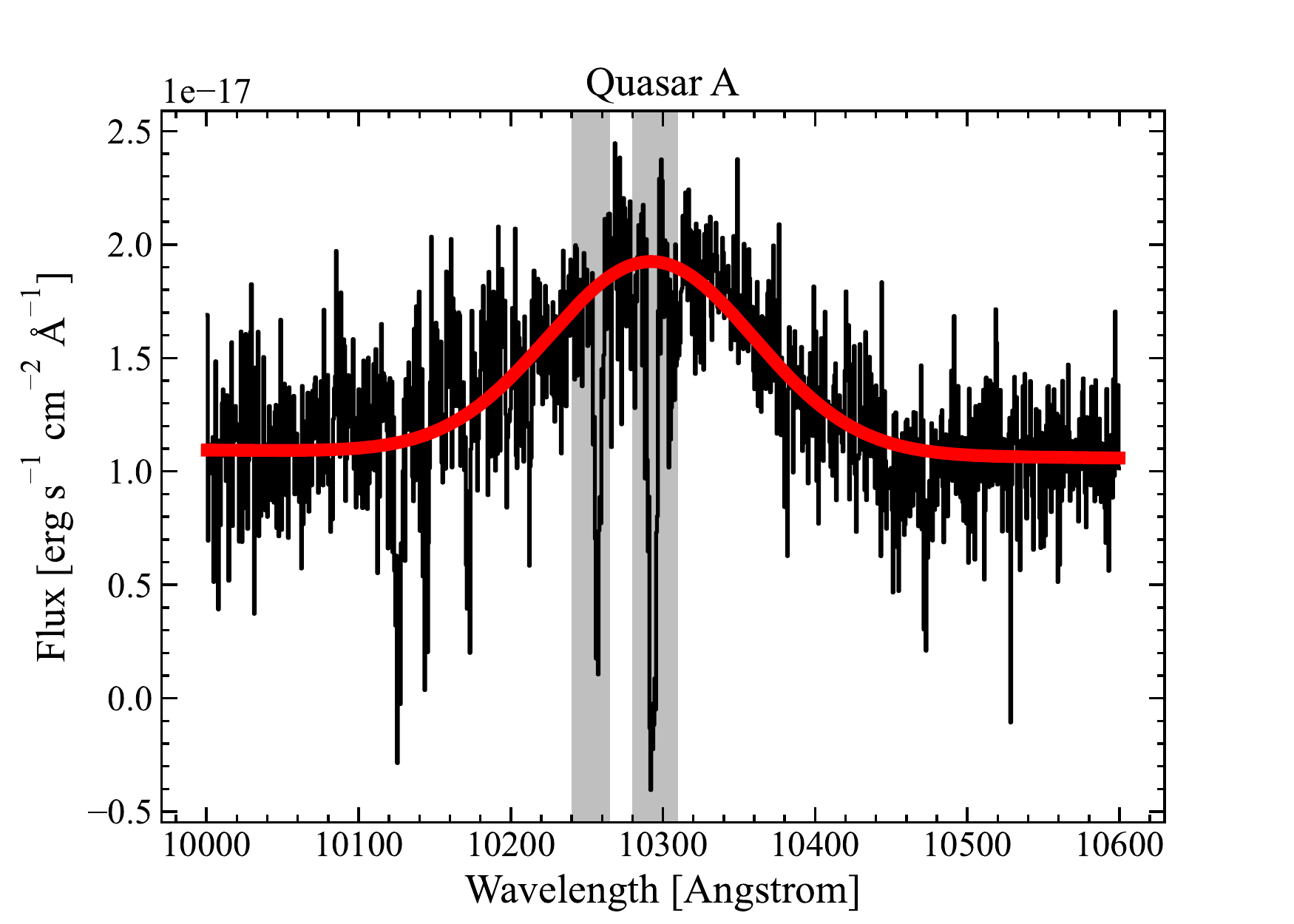}
    \includegraphics[width=0.98\columnwidth]{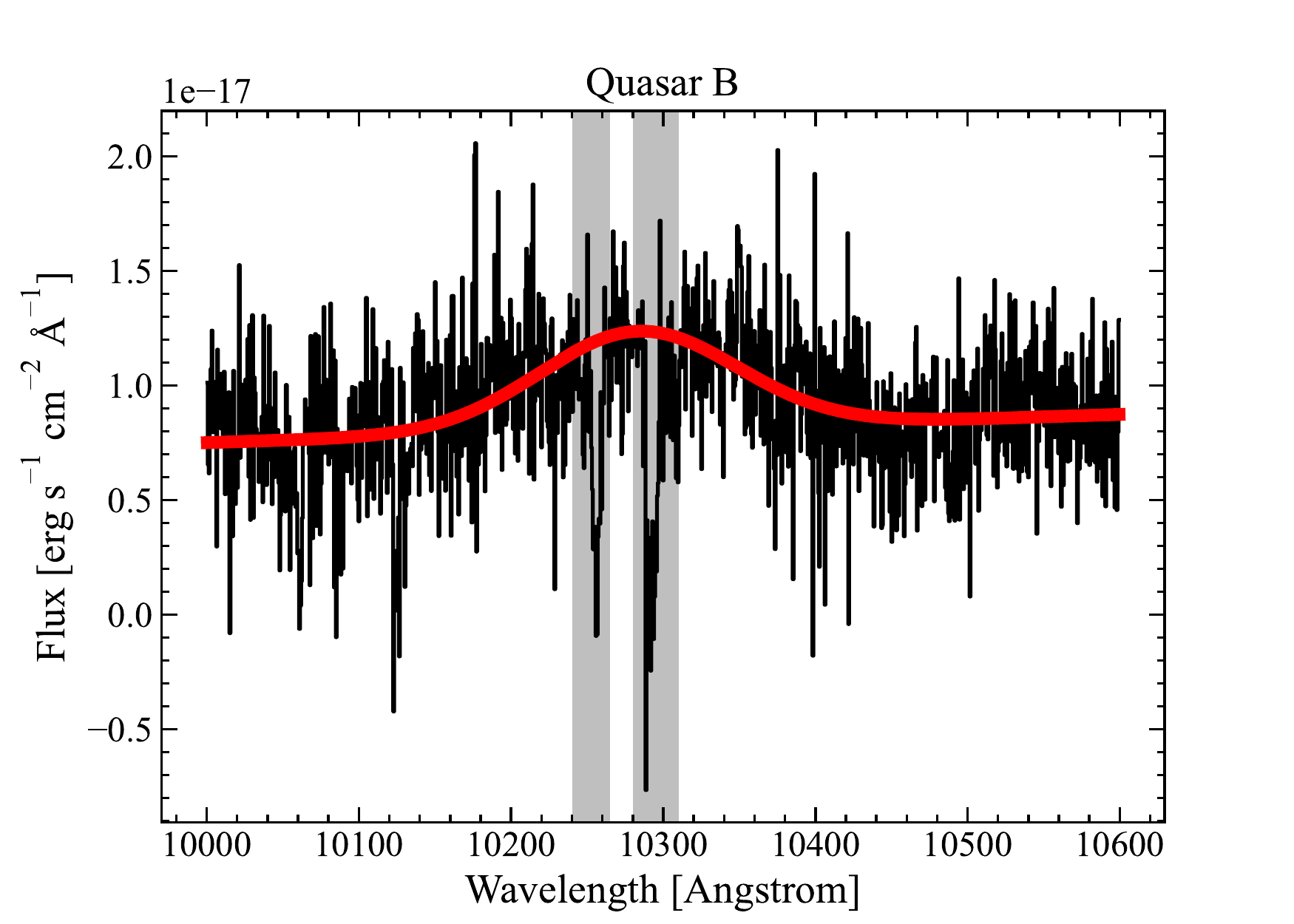}
    \caption{The {\civ} emission lines of the two quasars (black) and the best-fit model (red).
    We use a power-law continuum and a Gaussian profile to describe the emission lines. The absorption features 
    (gray dashed area) are masked during the fitting.}
    \label{fig:civ}
\end{figure}

Figure \ref{fig:civ} illustrates the {\civ} broad emission lines of the two objects, 
which are fitted as a power-law continuum plus a Gaussian profile.
Using the {\civ} line width and the empirical relation in \citet{vp06},
we estimate the SMBH masses to be $\log M_\text{BH}/M_\odot=8.60$ for quasar A and 
$\log M_\text{BH}/M_\odot=8.45$ for quasar B.
Although the {\mgii} line is a better SMBH mass indicator for high-redshift quasars, 
it falls in the wavelengths with strong sky emission lines at $z=5.66$.
The results are listed in Table \ref{tbl:properties}.

\begin{deluxetable}{ccc}
\tablenum{1}
\label{tbl:properties}
\tablecaption{Properties of J2037--4537}
\tablewidth{0pt}
\tablehead{\colhead{Object} & \colhead{A} & \colhead{B}}
\startdata
\hline
RA & 20:37:21.27 & -45:37:48.8\\
Dec & 20:37:21.26 & -45:37:47.5\\\hline
\multicolumn{3}{c}{Magnitudes\tablenotemark{1}}\\\hline
DES $g$ & $>25.1$ & $>25.1$ \\
DES $r$ & $23.09\pm0.04$ & $24.8\pm0.2$ \\
DES $i$ & $20.84\pm0.01$ & $21.80\pm0.02$ \\
DES $z$ & $20.17\pm0.01$ & $20.67\pm0.01$ \\
DES $Y$ & $20.07\pm 0.03$ & $20.39\pm0.05$ \\
F850LP & $20.10\pm0.01$ & $20.67\pm0.02$ \\\hline
\multicolumn{3}{c}{Luminosities}\\\hline
$M_\text{1450}$\tablenotemark{2} & $-26.42$ & $-25.99$\\
$\log L_\text{bol}(\text{erg s}^{-1})$ \tablenotemark{3}& 47.1 & 46.9 \\\hline
\multicolumn{3}{c}{{\civ} line}\\\hline
$z_\text{\civ}$ & $5.643\pm0.002$ & $5.636\pm0.003$ \\
FWHM $(\text{km s}^{-1})$& $1967\pm76$ & $1891\pm147$ \\
$\log M_\text{BH, \civ} (M_\odot)$ \tablenotemark{4}& 8.60 & 8.45\\\hline
\enddata
\tablenotetext{1}{All magnitudes are AB magnitudes. Detection limits are $5\sigma$.}
\tablenotetext{2}{The absolute magnitude at rest-frame 1450\AA.}
\tablenotetext{3}{Calculated based on $M_\text{1450}$ and the bolometric correction in \citet{runnoe12}.}
\tablenotetext{4}{The SMBH mass based on the relation in \citet{vp06}. The intrinsic scatter of the relation is 0.36 dex
which dominates the uncertainties.}
\end{deluxetable}

\subsection{HST Imaging} \label{subsec:hst}
\begin{figure}
    \centering
    \includegraphics[width=0.98\columnwidth]{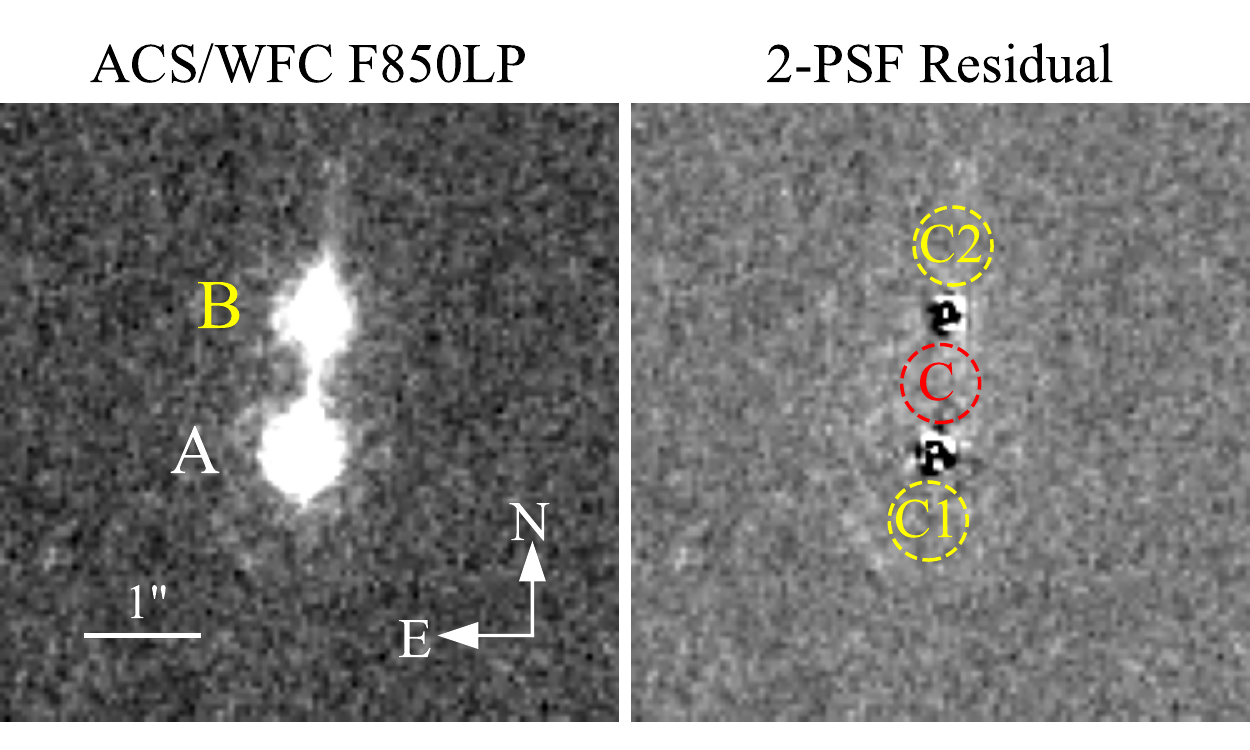}
    \caption{The {\em HST} ACS/WFC F850LP image of J2037--4537 (left) and the residual of the two-PSF model (right). There is no sign of a third object besides the two point sources.
    We measure the flux in the area C (the red circle, with a radius of $0\farcs35$) 
    to estimate the flux limit of the possible foreground lensing galaxy. 
    We also measure the fluxes in two additional areas, C1 and C2 (yellow circles), 
    which correct the contributions from the PSF-subtraction residual (see text for details). 
    Our estimation gives a $3\sigma$ flux limit of $m_\text{F850LP}>26.7$ for region C.
    }
    \label{fig:hst}
\end{figure}

Gravitational lensing can also generate two close quasar images at the same redshift. 
To fully explore the lensing hypothesis, 
we observed J2037--4537 using {\em HST} ACS/WFC in the F850LP filter, aiming at detecting or ruling out the foreground lensing galaxy (Program GO-16507).
The F850LP filter has a wavelength coverage of $0.8\mu\text{m}<\lambda<1.1\mu\text{m}$,
where the two objects show significantly different SEDs in the ground-based spectroscopy.
Figure \ref{fig:hst} presents the {\em HST} image and the two-PSF fitting residual.
J2037--4537 is well-described by the two-PSF model and there is no evidence for a third object in this field.

To further set a flux limit for the possible lensing galaxy,
we measure the flux in the region between the two point sources
(region C, marked by the red circle in the right panel of Figure \ref{fig:hst}).
Since PSF models of ACS/WFC usually have non-negligible errors in PSF wings \citep{jee07},
we measure two additional regions (C1 and C2, marked by the yellow circles) to correct this systematic uncertainty.
The yellow regions have the same size as the red one, and the distance from region C1 (C2) to object A (B)
equals to the distance from region C to object A (B).
The flux in region C1 (C2) thus estimates the contribution of the PSF wing residuals from object A (B) to the flux in region C.
We estimate the flux of the possible foreground galaxy as 
\begin{equation}
    F_\text{foreground} = F_\text{C} - F_\text{C1} - F_\text{C2}
\end{equation}
This gives a non-detection and rules out the existence of a foreground lensing galaxy
with $m_\text{F850LP}<26.7$ at $3\sigma$ level.
Assuming a typical redshift of $z_l\sim1$ for the lensing galaxy 
\citep[e.g.,][]{hilbert08, wyithe11},
this limit gives an absolute magnitude of $M\gtrsim-17.6$, which is about four magnitudes fainter than
the breaking magnitude of the galaxy luminosity function at $z\sim1$
\citep[$M_*\sim-21.5$; e.g.,][]{faber07}.

\subsection{Testing the Strong-Lensing Hypothesis}

The distinct SEDs of the two quasar images and the non-detection of the lensing galaxy
strongly disfavor the lensing hypothesis. Although microlensing and/or differential reddening 
can lead to different spectral features in lensed quasar images \citep[e.g.,][]{sluse12}, 
it is difficult for these effects to explain the observational features of J2037--4537.
If object A and B are lensed images of the same quasar, the extinction of 
image B must be much stronger than A given their colors. At long wavelengths
$(\lambda_\text{obs}\gtrsim1.5\micron)$, image B is brighter than image A,
meaning that image A is intrinsically fainter. Meanwhile, 
in most galaxy-scale lensing systems, the fainter image is closer to the deflector 
\citep[e.g.,][]{mason15} and should have a stronger extinction.
This conflicts with the fact that image B is redder than A,
suggesting that the lensing scenario is unlikely.



In principle, a deflector galaxy with highly irregular mass profiles and/or dust distributions
might be able to generate a lensing system like J2037--4537.
We thus report J2037--4537 as a high-confidence close quasar pair candidate.
The definitive test of the lensing scenario
can be achieved with the Atacama Large Millimeter/submillimeter Array (ALMA).
ALMA will measure the {\cii} redshifts of the two quasar images to an accuracy of $<0.001$
\citep[e.g.,][]{decarli18}, and will reveal the lensed arcs of the quasar host galaxy. 
In the following discussions, however, we will treat J2037--4537 as a physical quasar pair.

\section{Discussion} \label{sec:discussion}

\subsection{The Pair Fraction of High-Redshift Quasars}

J2037--4537 has a projected separation of $1\farcs24$ (7.3 kpc at $z=5.66$).
Kiloparsec-scale quasar pairs are rare especially at high redshift,
due to the quick decline of the number density of high-redshift quasars.
\citet{dr19} summarize a number of predictions of the active galactic nuclei (AGNs) pair fractions
\citep[e.g.,][]{steinborn16,volonteri16,rg19}.
These studies use large-volume cosmological simulations like 
Magneticum\footnote{http://www.magneticum.org/index.html},
EAGLES \citep{eagle}, and Horizon-AGN \citep{horizon},
which model the SMBH growth and AGN activities using analytical relations. 
These studies then count close SMBH pairs that have bolometric luminosities beyond a certain threshold.
Specifically, \citet{dr19} consider AGNs with bolometric luminosity $L_\text{bol}>10^{43}\text{~erg s}^{-1}$,
and count AGN pairs with a separation of $r<30$ proper kpc (pkpc).
The predicted pair fraction, $f_\text{pair}(r<30\text{~pkpc})$, is about $1\%$ with little redshift evolution.


J2037--4537 allows us to constrain the quasar pair fraction at high redshift for the first time. 
This is done by estimating the expected number of quasars we can find in our survey.
J2037--4537 is discovered serendipitously in our survey for
high-redshift lensed quasars in the DES field, which covers $\sim5000\text{~deg}^2$.
We identify J2037--4537 as a lens candidate with quasar-dominated flux,
for which we request a non-detection in DES$-g$, a DES$-z$ magnitude of $m_z<21$,
and a projected separation of $\Delta d<3\farcs0$.
The $g-$ and $z-$band flux cuts restrict our targets to $5\lesssim z\lesssim6$ quasars \citep[e.g.,][]{wang16,yang17,yang19}.
As the survey is not completed yet, we simply assume a selection function of one
for $5<z<6$ quasars,
and provide the lower limit of the quasar pair fraction.

We use SIMQSO \citep{mcgreer13} to generate a mock catalog of quasars 
in the DES field at $5<z<6$ that are brighter than $m_z=21$.
SIMQSO is a Python-based package which generates mock catalogs of quasars with simulated spectra, 
and has been shown to accurately reproduce the observed colors and magnitudes of quasars \citep[e.g.,][]{ross13}. 
We use a broken power-law to parameterize the quasar luminosity function (QLF):
\begin{equation}
    \Phi = \frac{\Phi^*}{10^{0.4(M-M^*)(\alpha+1)}+10^{0.4(M-M^*)(\beta+1)}}
\end{equation}
We adopt the $z=5$ QLF from \citet{kim20} and the $z=6$ QLF from \citet{matsuoka18},
and linearly interpolate the QLF parameters $(\log\Phi^*, M^*, \alpha, \beta)$ for other redshifts at $5<z<6$.
SIMQSO suggests that there are $N_\text{all}=485$ quasars brighter than $m_z=21$ at $5<z<6$ in the DES field.

J2037--4537 directly constrains the number of quasar pairs at $z>5$ with $0\farcs9<\Delta d<3\farcs0$.
The lower boundary of the projected separation corresponds to the spatial resolution of the DES survey.
To avoid confusion, we always use $r$ to denote the physical separation of a quasar pair,
and use $\Delta d$ to denote the projected separation.
By converting the angular separations to proper distances, we find
$f_\text{pair}[5.3<\Delta d~(\text{pkpc})<17.7]\ge1/485=0.2\%$.
As a comparison, the expected number of quasar pairs is $7\times10^{-6}$ if quasars are randomly distributed.
We estimate this number by assigning random positions to the quasars in the mock catalog, 
and counting pairs of mock quasars with $0\farcs9<\Delta d<3\farcs0$ and $\Delta v<2000\text{ km s}^{-1}$,
following \citet{hennawi06}.
The clustering signal for $z>5$ quasars is thus $W_p(0\farcs9<\Delta d<3\farcs0)\sim10^5$.
\citet{hennawi06} measure the quasar correlation function at $1\lesssim z\lesssim 3$ and 
$\Delta d\sim 0.1\text{ pMpc}$. Extrapolating the correlation function 
in \citet{hennawi06} to small scales gives a clustering signal of $W_p\sim10^3$.
This comparison indicates that close quasar pairs are regulated by processes 
(likely galaxy mergers) that are different from quasar clustering at larger scale.

In order to make direct comparisons to the simulated pair fractions,
we convert $f_\text{pair}(0\farcs9<\Delta d<3\farcs0)$ to $f_\text{pair}(r<30\text{~pkpc})$ as follows.
We assume that the small-scale quasar correlation function can be described by a power-law,
i.e., $\xi (r)= (r/R_0)^{-\gamma}$. Following the analysis in  \citet{hennawi06} and \citet{mcgreer16},
the expected number of quasar pairs with separation $r<R$ can be calculated as
\begin{equation} \label{eq:npair1}
    N_\text{pair}(r<R) = \frac{1}{2}\sum_j^{N_\text{all}}\int^R_0 n(z_j)\times 4\pi r^2 [1+\xi(r)]dr
\end{equation}
where the sum goes through all the mock quasars and $n(z_j)$ is the number density of quasars at redshift $z_j$
that have $m_z<21$. We use the quasar spectrum template from \citet{vdb01} to 
convert the absolute magnitude $M_{1450}$ to the $z-$band apparent magnitude.
Similarly, the expected number of quasar pairs with projected separation $d_\text{min}<\Delta d<d_\text{max}$ is
\begin{multline} \label{eq:npair2}
    N_\text{pair}(d_\text{min}<\Delta d<d_\text{max}) =\\ \frac{1}{2}\sum_j^{N_\text{all}}\int^{+\frac{v_\text{max}}{aH(z_j)}}_{-\frac{v_\text{max}}{aH(z_j)}}dy\int^{d_\text{max}}_{d_\text{min}}dx~n(z_j) \times 2\pi x [1+\xi(\sqrt{y^2+x^2})]
\end{multline}
where $a=(1+z)^{-1}$ is the scale factor, $H(z_j)$ is the Hubble constant at redshift $z_j$,
and $v_\text{max}=2000\text{ km s}^{-1}$ is the maximum velocity difference of the quasar pair.
Note that all the quantities in Equation \ref{eq:npair1} and \ref{eq:npair2} are in comoving units.
We adopt a fiducial power-law index of $\gamma=2$, which gives a good description of quasar pairs
down to a separation of $\lesssim100$ pkpc \citep[e.g.,][]{shen10}. At small separations, we have $\xi(r)\gg1$,
and $\gamma=2$ gives a flat distribution of $\dv{N_\text{pair}(r<R)}{R}$, in agreement with the 
simulated AGN pairs in \citet{rg19}.

\begin{figure}
    \centering
    \includegraphics[width=0.98\columnwidth]{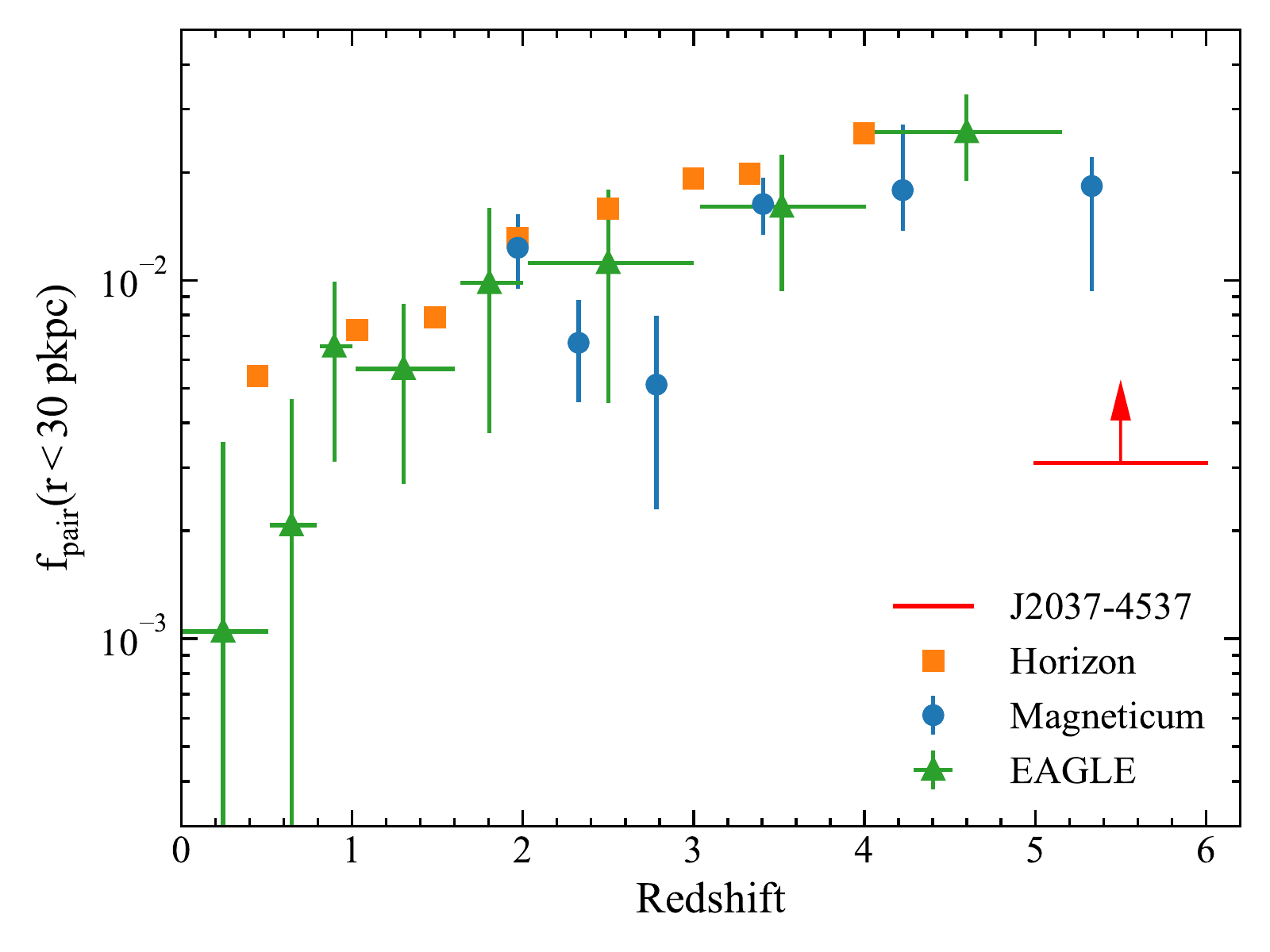}
    \caption{The pair fraction of quasars, $f_\text{pair}(r<30\text{ pkpc})$. 
    The red line marks the lower limit set by J2037--4537. We also include the predictions of cosmological simulations,
    which are adapted from the Figure 6 in \citet{dr19}. 
    The simulated AGN samples have $L_\text{bol}>10^{43}\text{erg s}^{-1}$,
    and include both obscured and unobscured AGNs.
    The observed lower limit is several times lower than the simulated sample.
    Note that the two quasars in  J2037--4537 have 
    $L_\text{bol}\sim10^{47}\text{erg s}^{-1}$ (Table \ref{tbl:properties}).
    Observations at lower redshift indicate that the pair fraction decreases with AGN luminosity,
    which could explain the difference between the simulations and the lower limit set by J2037--4537
    (see text for details).}
    \label{fig:pairfrac}
\end{figure}

By applying $N_\text{pair}(0\farcs9<\Delta d<3\farcs0)>1$ to the above equations,
we get $R_0>254h^{-1}$ comoving Mpc (cMpc), $N_\text{pair}(r<30\text{ pkpc})>1.5$ and $f_\text{pair}(r<30\text{ pkpc})>0.3\%$. 
Figure \ref{fig:pairfrac} shows the comparison between our results and the predictions from the cosmological simulations. 
The lower limit is several times lower than the predicted pair fraction.
Note that the simulations count both obscured and unobscured AGNs.
Since our survey only targets unobscured type-I quasars, 
we expect that the observed quasar pair fraction in our survey should be lower than 
 the simulated AGN pair fractions.
In addition, the AGN pair fraction depends on the luminosity and the SMBH mass, 
as discussed in \citet{dr19}.
\citet{silverman20} give an observed pair fraction of $\sim0.26\%$ for $z\lesssim4$ quasars,
and their sample has a luminosity cut $(\log L_\text{bol}>45.3)$ that is comparable to our survey.
\citet{silverman20} show that the observed fraction is consistent with a luminosity-matched sample in the cosmological simulations,
which indicates that the pair fraction might be lower for luminous AGNs.
Current cosmological simulations do not have sufficient volumes to produce a quasar pair like J2037--4537 
at $z>5$. Future simulations with larger volumes will enable direct comparisons with observations 
and provide unique constraints on the SMBH evolution models at high redshift.

\subsection{The Triggering Mechanism of the Quasar Pair}

The tangential separation between the two quasars in J2037--4537 is $7.3\text{ kpc}$,
and the spectra suggest a redshift difference of $\Delta z\lesssim0.01$.
As such, this quasar pair must reside in a galaxy merger.
Although current data cannot rule out the possibility that the two quasars are triggered independently and coincidentally
before the merging event, we argue that J2037--4537 is likely triggered by the galaxy merger.
This is because the correlation length indicated by J2037--4537  $(R_0>254h^{-1} \text{~cMpc})$
is much larger than the correlation length of $z\sim5$ quasars 
\citep[$R_0\sim20h^{-1}$ cMpc, e.g.,][]{mcgreer16}. 
The significant difference in $R_0$ at large and small scales
indicates that close quasar pairs and quasar pairs with separation $\gtrsim100\text{ pkpc}$
might have different origins. The former are more likely results of galaxy mergers, 
while the latter are related to structure formation at sub-Mpc scale.

The observational features of J2037--4537 is consistent 
with merger-triggered quasar pairs in hydrodynamical simulations.
\citet{capelo17} explore the factors that control the emergence of AGN pairs in galaxy mergers,
and conclude that close quasar pairs usually appear in galaxy mergers with SMBH mass ratios close to one
and separations less than 10 pkpc.
It is thus plausible that the two quasars are triggered by the merging process.
Future observations (likely with ALMA or the {\em James Webb Space Telescope})
will reveal more details of the gas kinematics and will allow investigating how the gas feeds the SMBHs
in this system.
Besides, we notice that the two quasars show similar profiles for
a few broad emission lines (e.g., the Ly$\alpha$, {\nv}, {\oi} and {\siiv} lines).
If J2037--4537 is a physical quasar pair, 
these similarities might be causally connected to the physical association of the two quasars
and encode critical information about how the quasar activities are triggered.

\section{Summary} \label{sec:sum}
We report the discovery of a close quasar pair candidate
 at $z=5.66$, J2037--4537, which has a projected separation 
of $1\farcs24$ $(7.3\text{~kpc})$. 
The ground-based spectroscopy shows that the resolved two objects are two quasars at the same redshift, 
and the high-resolution {\em HST} imaging 
does not detect the foreground lensing galaxy.
Given the features in the spectra and the high-resolution images,
it is highly unlikely that J2037--4537 is a lensed quasar.
J2037--4537 sets the first constraint on the pair fraction of quasars at $5<z<6$, $f_\text{pair}(r<30\text{~pkpc})>0.3\%$. 
Future observations of the gas kinematics in J2037--4537 will 
reveal more information about the triggering mechanism of the quasar pair.

\acknowledgments
We thank the referee and the editor for their valuable comments.
This research is based on observations made with the NASA/ESA Hubble Space Telescope obtained from the Space Telescope Science Institute, which is operated by the Association of Universities for Research in Astronomy, Inc., under NASA contract NAS 5–26555. These observations are associated with program GO-16507.
Some of the data presented in this paper were obtained from the Mikulski Archive for Space Telescopes (MAST) at the Space Telescope Science Institute. The specific observations analyzed can be accessed via \dataset[10.17909/t9-gw19-6w51]{https://doi.org/10.17909/t9-gw19-6w51}.
MY, XF and JY acknowledges supports by NSF grants AST 19-08284.
MY and XF ackowledges supports by HST-GO-16507 grant from the Space Telescope Science Institute. 
FW thanks the support provided by NASA through the NASA Hubble Fellowship grant \#HST-HF2-51448.001-A awarded by the Space Telescope Science Institute, which is operated by the Association of Universities for Research in Astronomy, Incorporated, under NASA contract NAS5-26555.
\facilities {\em HST}, Magellan/Clay, Magellan/Baade

\bibliography{sample631}{}
\bibliographystyle{aasjournal}



\end{document}